\documentclass[twocolumn,aps,prl,epsfig,graphics,showpacs]{revtex4}%
\usepackage{graphicx}
\usepackage{amsmath}
\usepackage{amsfonts}
\usepackage{bbm}
\usepackage{amssymb}
\begin{document}

\title{Strong magnetic coupling between an electronic spin qubit and a mechanical resonator}

\author{P. Rabl$^{1,2}$, P. Cappellaro$^{1,2}$,  M. V. Gurudev Dutt$^{3}$, L. Jiang$^2$, J. R. Maze$^2$,  and M. D. Lukin$^{1,2}$ }

\affiliation{$^1$ ITAMP, Harvard-Smithsonian Center for Astrophysics, Cambridge, MA 02138, USA}
\affiliation{$^2$ Department of Physics, Harvard University, Cambridge, MA 02138, USA}
\affiliation{$^3$ Department of Physics and Astronomy, University of Pittsburgh, Pittsburgh, PA 15260, USA}


\begin{abstract}
We describe a technique that enables a strong, coherent coupling between a single electronic spin qubit associated with a nitrogen-vacancy impurity in diamond and the quantized motion of a magnetized nano-mechanical resonator tip. This coupling is achieved via careful preparation of dressed spin states which are highly sensitive to the motion of the resonator but insensitive to perturbations from the nuclear spin bath. 
In combination with optical pumping techniques, the coherent exchange between spin and motional excitations enables ground state cooling and the controlled generation of arbitrary quantum superpositions of resonator states. Optical spin readout techniques provide a general measurement toolbox for the resonator with quantum limited precision.
\end{abstract}



\maketitle

Techniques for cooling and quantum manipulation of motional states of nano-mechanical resonators are now actively explored. Work in this field
is motivated by ideas from quantum information science~\cite{EisertPRL2004,VitaliPRL2007}, testing
quantum mechanics for macroscopic objects~\cite{ArmourPRL2002,MarshallPRL2003} and potential
applications in nano-scale sensing~\cite{SidlesRMP1995,MaminNatureNano2007}. Approaches based on mechanical resonators coupled to optical cavities~\cite{OpticalCavExp}, superconducting devices~\cite{NaikNature2006,Teufel} or cold atoms~\cite{TreutleinPRL2007} are
presently being investigated in experiments.

In this paper we describe a technique that enables a coherent coupling between the quantized motion of a mechanical resonator and an isolated spin qubit. Specifically, we focus on
the electronic spin associated with a nitrogen-vacancy (NV) impurity in diamond~\cite{Wachtrup2006} which
can be optically polarized and detected, and exhibits excellent coherence properties even at room temperature~\cite{NVCoherentControl}. Since its precession frequency depends on external magnetic fields via the Zeeman
effect, single spins can be used as magnetic sensors operating at nanometer scales~\cite{NVMagnetometerTH,NVMagnetometerEXP}.

\begin{figure}
\begin{centering}
\includegraphics[width=0.45\textwidth]{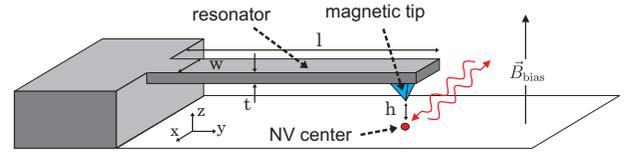}
\caption{A magnetic tip attached to the end of a nano-mechanical resonator of dimensions $(l,w,t)$ is positioned at a distance $h\sim 25$ nm above a single NV center, thereby creating a strong coupling between the electronic spin of the defect center and the motion of the resonator. Microwave and laser fields are used to manipulate and measure the spin states.  } \label{fig:Setup}
\end{centering}
\end{figure}

The essential idea of the present
work can be understood by considering a prototype system shown in Fig.~\ref{fig:Setup}. Here a single spin is used to sense the motion of the magnetized resonator tip, that is separated from the spin by an average distance $h$ and oscillates at frequency $\omega_r$. These oscillations produce a time-varying magnetic field
that causes Zeeman shifts of the spin qubit. Specifically, the shift corresponding to a single quantum of motion is
$\lambda = g_s \mu_B G_m a_0$, where $g_s\simeq 2$, $\mu_B$ is the Bohr magneton, $G_m$ the magnetic field gradient and $a_0=\sqrt{\hbar/2m\omega_r}$ the amplitude of zero-point fluctuations for a resonator of mass $m$. For realistic conditions, $h \approx 25$ nm, $\omega_r/2\pi \approx 5$ MHz,  $a_0\approx 5\times 10^{-13 }$ m and $G_m\approx 10^7$ T/m we find that
$\lambda/2\pi$ can approach 100 kHz. Such a large shift can be easily measured within a fraction of a millisecond by detecting the electronic spin state~\cite{NVMagnetometerEXP}. More
importantly, the coupling constant $\lambda$ can considerably exceed both the electronic spin coherence time ($T_2 \sim 1$ ms) and the intrinsic damping rate, $\kappa = \omega_r/Q$, of high-$Q$ mechanical resonators. In this regime, the spin becomes strongly
coupled to mechanical motion in direct analogy to strong coupling of cavity quantum electrodynamics (QED).

Before proceeding we note that coupling of mechanical motion to several types of matter qubits, ranging from Cooper pair boxes to trapped atoms, has been considered previously~\cite{ArmourPRL2002,TreutleinPRL2007,WilsonRaePRL2004}. The distinguishing feature of the present approach is that working
at nano-scale dimensions allows us to combine a well isolated spin qubit with a large interaction strength, thus enabling the strong coupling regime.  In what follows we describe how this regime can be accessed in the presence of fast dephasing ($T_2^* \sim 1 \,\mu$s) of the electronic spin due to interactions with the nuclear spin bath by using an appropriate dressed spin basis.
We then show how it can be applied to
cooling and quantum manipulation of mechanical motion.

%

In the setup shown in Fig.~\ref{fig:Setup} the nano-mechanical resonator is described by the Hamiltonian $H_r=\hbar \omega_r a^\dag a$ with $\omega_r$ the frequency of the fundamental bending mode and $a$, $a^\dag$ the corresponding annihilation and creation operators. Motion of the magnetic tip produces a field $|\vec B_{\rm tip}|\simeq G_m \hat z$, which is proportional to the position operator $\hat z = a_0 (a+a^\dag)$ and results in a Hamiltonian
\begin{equation}\label{eq:Hsys}
H_S= H_{NV}+\hbar \omega_r a^\dag a + \hbar\lambda (a+a^\dag)S_z\,.
\end{equation}
Here $H_{NV}$ describes the dynamics of the driven electronic spin  and $S_z$ is the z-component of the spin operator which we here assume to be aligned with the NV symmetry axis.



\begin{figure}
\begin{centering}
\includegraphics[width=0.48\textwidth]{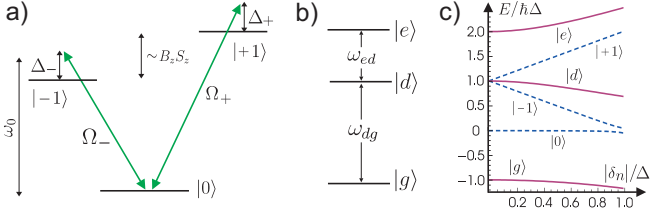}
\caption{(a) Level diagram of the driven NV center in the electronic ground state. (b) Dressed spin basis states for $\Omega\sim |\Delta|$ and $\Delta <0$.   (c) Energies of dressed states in the presence of external perturbations of the form $H_{\rm nuc}= \hbar \delta_n S_z$ for $\Omega/|\Delta|=0.1$ (dashed) and $\Omega/|\Delta|=1$ (solid).} \label{fig:spinstates}
\end{centering}
\end{figure}

The electronic ground state of the NV center is an $S=1$ spin triplet and we label states by $|m_s\rangle$, $m_s=0,\pm 1 $. 
Spin states with different values of $|m_s|$ are separated by a zero field splitting of $\omega_0/2\pi \simeq 2.88$ GHz. For moderate applied magnetic fields~\cite{Note1}, $|\mu_B \vec B|\ll \hbar \omega_0$, static and low-frequency components of magnetic fields cause Zeeman shifts of states $|\pm 1\rangle$ while microwave (mw) fields of appropriate polarization drive Rabi oscillations between $|0\rangle$ and the exited states $|\pm 1\rangle$ as shown in Fig.~\ref{fig:spinstates}(a). In a frame rotating with the frequencies of the mw fields,
\begin{equation}\label{eq:HNV1}
H_{NV}=\sum_{i=\pm1} -\hbar \Delta_i |i\rangle\langle i| +\frac{\hbar\Omega_i}{2} \left(|0\rangle \langle i | + |i\rangle\langle 0| \right),
\end{equation}
where $\Delta_\pm$ and $\Omega_\pm$ denote the detunings and the Rabi frequencies of the two microwave transitions. For simplicity we restrict the following discussion to symmetric conditions, i.e.  $\Delta_i\equiv \Delta$, $\Omega_i\equiv \Omega$ (e.g. using a single mw-field and $B\rightarrow 0$).   Hamiltonian~\eqref{eq:HNV1} then couples the state $|0\rangle$ to a `bright' superposition of excited states $|b\rangle =(|\!-\!1\rangle +|\!+\!1\rangle)/\sqrt{2}$, while the `dark' superposition $|d\rangle =(|\!-\!1\rangle-|\!+\!1\rangle)/\sqrt{2}$ remains decoupled. The resulting eigenbasis of $H_{NV}$ is therefore given by $|d\rangle$ and the two dressed states $|g\rangle=\cos(\theta) |0\rangle-\sin(\theta)|b\rangle$ and $|e\rangle = \cos(\theta)|b\rangle + \sin(\theta) |0\rangle$, with $\tan(2\theta)=-\sqrt{2}\Omega/\Delta$. Corresponding eigen-frequencies are $\omega_d= -\Delta$ and $\omega_{e/g}= (-\Delta \!\pm \! \sqrt{\Delta^2\!+\!2\Omega^2})/2$. We will mainly focus on the regime $\Delta <0 $ where $|g\rangle$ is the lowest energy state (Fig.~\ref{fig:spinstates}(b)).

To achieve a resonant coupling, values for $\Omega$ and $|\Delta|$ can now be adjusted such that transition frequencies  between dressed states, e.g. $\omega_{dg}=\omega_d-\omega_g$, become comparable with the oscillator frequency $\omega_r$.
Rewriting Hamiltonian~\eqref{eq:Hsys} in terms of  $|g\rangle$, $|d\rangle$ and $|e\rangle$ we obtain
 \begin{equation}\label{eq:HCavityQED}
\begin{split}
H_S= &\,\,\hbar \omega_r a^\dag a + \hbar \omega_{eg}|e\rangle\langle e| + \hbar \omega_{dg} |d\rangle\langle d| \\
 & +\hbar (\lambda_g|g\rangle\langle d | + \lambda_e |d\rangle\langle e | + {\rm H.c.} ) (a+a^\dag) \,,
\end{split}
\end{equation}
where $\lambda_g= -\lambda \sin(\theta)$ and $\lambda_e= \lambda \cos(\theta)$. Under resonance conditions, $\omega_r\approx |\omega_{gd}|$ Hamiltonian~\eqref{eq:HCavityQED} reduces to the well-known Jaynes Cummings model (JCM), and describes coherent oscillations between states $|n\rangle|g\rangle$ and $|n-1\rangle|d\rangle$ where $|n\rangle$ denotes a phonon number state.
To observe the coherent dynamics associated with this model the vacuum Rabi frequency $\lambda_{g}$ must be compared to the motional decoherence rate $\gamma_r$ and random shifts of the transition frequency, $\Delta \omega_{dg}$, due to hyperfine interactions with the nuclear spin bath. The condition $\lambda_g \gg \gamma_r,\Delta \omega_{dg}$ then defines the strong coupling regime. While in principle $\gamma_r\equiv \kappa$ at zero temperature, we identify below $\gamma_r\equiv k_BT/\hbar Q$ as the relevant decoherence rate for experimentally accessible temperatures $T$. Interactions with the nuclear spin bath are characterized by a typical strength $\delta_n \sim 1/T_2^* \sim 1$ MHz which exceeds $\lambda$.

To show how the strong coupling regime can be achieved,
we now study the driven NV center in the presence of hyperfine interactions, $H_{\rm nuc}= g_s\mu_B B_{n,z}(t)S_z$, where $\vec B_n(t)$ is the effective magnetic field associated with the nuclear spin bath. $\vec B_n(t)$ is quasi-static on the timescales of interest~\cite{Note1} but has a random magnitude on the order $|B_{n,z}|=\hbar \delta_n/(g_s\mu_B)$. In Fig.~\ref{fig:spinstates}(c) we plot dressed state energies of $H_{NV}'=H_{NV}+H_{\rm nuc}$ as a function of $\delta_n$. For $\Omega\rightarrow 0$ we recover the linear Zeeman shift for the  bare spin states $|\pm 1\rangle$. In this regime large random shifts of $\omega_{dg}$ would prevent resonant interactions between the spin and the resonator mode. However, for $\Omega \sim |\Delta|$ where the operator $S_z$ has only off-diagonal matrix elements in the dressed state basis, perturbations are suppressed for $\Omega \gg |\delta_n|$. In particular, the quadratic shift of the transition frequency $\omega_{dg}$ is given by
\begin{equation}\label{eq:SecondOrder}
\Delta \omega^{(2)}_{dg}= \delta^2_n/\omega_{gd} \left(1- \tan^{-2}(\theta) +  \sin^2(\theta)  \right).
\end{equation}
We find that in the present three level configuration not only we can eliminate linear shifts of the transition frequency, but at a particular value of $\theta\approx \theta_0\simeq 0.22\pi$ even quadratic corrections vanish. At this `sweet spot' the remaining frequency shift is $\Delta \omega_{gd}^{(4)}\approx 0.6 \times\delta_n^4/\omega_{gd}^3$. In other words,  operation at the sweet spot allows us to suppress the effect of $H_{\rm nuc}$ by several orders of magnitude and treat remaining corrections as a small perturbation.

\emph{Example.}
As an example we consider a Si-cantilever~\cite{SidlesRMP1995} of dimensions $(l,w,t)=(3,0.05,0.05)\,\mu m$ with a fundamental frequency of $\omega_r/2\pi\approx 7$ MHz and $a_0\approx 5\times 10^{-13}$ m. A magnetic tip of size $\sim 100$ nm and homogeneous magnetization $M\approx 2.3\times  10^6  \,{\rm T}/\mu_0$~\cite{MaminNatureNano2007} produces a magnetic gradient of $G_m\approx 7.8\times 10^6$ T/m at a distance $h\approx 25$ nm away from the tip and results in a coupling strength  $\lambda/2\pi\approx 115$ kHz. For a temperature of $T=100$ mK and  Q values of $10^5$ the heating rate is $\gamma_r/2\pi\approx 20$ kHz. Operating close to the sweet spot $\theta\approx \theta_0$ and assuming $\delta_n\simeq 1$ MHz we obtain $(\lambda_g,\gamma_r, \Delta \omega_{gd})=2\pi\times (70,20, 2)$ kHz.  Hence, this combination enables us to access the strong coupling regime of Hamiltonian~\eqref{eq:HCavityQED}.  Higher values of $Q \sim 10^6$ and/or a reduction of the system dimensions to $h\sim 10$ nm would allow further improvements. The strong coupling regime can then  be reached
even at temperatures up to a few K.



\emph{Applications.} Hamiltonian~\eqref{eq:HCavityQED} allows a coherent
transfer of quantum states between the spin and the resonator mode, which in combination with optical pumping and readout techniques for spin states~\cite{NVCoherentControl} provides the basic ingredients for the generation and detection of various non-classical states of the mechanical resonator. Here we  discuss in more detail an optical cooling scheme to prepare the resonator close to the quantum ground state and a general strategy for the generation of arbitrary superpositions of resonator states.

\begin{figure}
\begin{centering}
\includegraphics[width=0.48\textwidth]{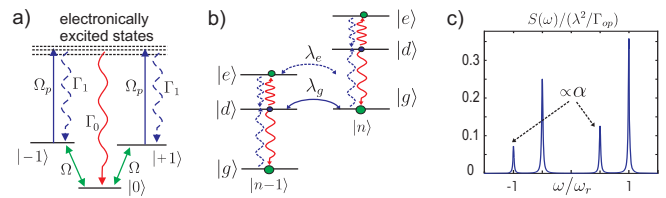}
\caption{(a) Optical pumping of spin states. (b) Level diagram of Hamiltonian $H_S$, Eq.~\eqref{eq:HCavityQED}. Wavy lines indicate optical pumping processes into and out of state $|d\rangle$ for $\alpha=0$ (solid) and $\alpha>0$ (solid and dashed).  (c) Excitation spectrum $S(\omega)$, Eq.~\eqref{eq:Spectrum} for $\omega_{dg}=\omega_r$, $\omega_{ed}\simeq 0.53\,\omega_r$, $\Gamma_{op}/\omega_r=0.01$ and $\alpha=1$. } \label{fig:Cooling}
\end{centering}
\end{figure}

Cooling and state preparation techniques rely on a controlled dissipation of energy 
which in the present setting can be achieved by optical pumping of spin states as shown in Fig.~\ref{fig:Cooling} a). A laser excites spin states $|\pm 1\rangle$ into higher electronic levels from where they decay with a rate $\Gamma_1$ back to the same spin state, or with a rate $\Gamma_0$ to state $|0\rangle$. Projected on the electronic ground state we can characterize the pumping process by a tunable pumping rate $\Gamma_{op}(t)=\Omega_p^2(t)\Gamma_0/(\Gamma_1\!+\!\Gamma_0)^2$ between states $|\pm 1\rangle$ and $|0\rangle$ and the branching ratio $\alpha=\Gamma_1/\Gamma_0$. While off resonant excitations at room temperature yield $\alpha\sim 1$, the ideal limit $\alpha \rightarrow 0$ can be reached using resonant excitations of appropriately chosen transitions at lower temperatures~\cite{OpticalTransitions}. Including mechanical dissipation of the resonator mode
the evolution of the system density operator $\rho(t)$ is described by the master equation
\begin{equation}
\begin{split}\label{eq:MasterEquation}
\dot \rho(t) = \,&i[\rho(t),H_S] + \kappa\left( (N_{th}\!+\!1) \mathcal{D}[a] +  N_{th} \mathcal{D}[a^\dag]\right)\rho\\
& + \Gamma_{op}(t) \sum_{i=\pm 1} \left( \mathcal{D}[|0\rangle\langle i|] + \alpha \, \mathcal{D}[|i\rangle \langle i| ] \right)\rho\,.
\end{split}
\end{equation}
Here $\mathcal{D}[\hat c]\rho := \left( 2\hat c \rho \hat c^\dag - \hat c^\dag \hat c \rho - \,\rho \hat c^\dag \hat c\right)/2$ and $N_{th}=[\exp(\hbar \omega_r/k_BT)-1]^{-1}$ is the thermal equilibrium occupation number for a support temperature $T$.

To remove thermal excitations we first study cw-cooling of the resonator mode. Assuming $\Gamma_{op}(t)\equiv \Gamma_{op} \gg \lambda$ we eliminate the fast dynamics of the spin degrees of freedom~\cite{CiracPRA1992} and study the effective evolution of the mean occupation number $\langle n\rangle (t)= {\rm Tr}\{ \rho(t) a^\dag a\}$. The resulting equation is of the form  $\langle \dot n\rangle= -W(\langle n\rangle -\langle n\rangle_0)$,
with a total cooling rate $W=W_{op} + \kappa$ and a final occupation number $\langle n\rangle_0=(\kappa N_{th} +A_{op}^+)/W$.
Here the optical cooling and heating rates, $W_{op}=S(\omega_r)-S(-\omega_r)$ and $A_{op}^+=S(-\omega_r)$, are determined by the  fluctuation spectrum
\begin{equation}\label{eq:Spectrum}
S(\omega)= 2 \lambda^2\, {\rm Re} \int_0^\infty d\tau \, \langle S_z(\tau)S_z\rangle_0 \, e^{i\omega\tau}\,,
\end{equation}
which describes the ability of the spin to absorb ($\omega >0$) or emit ($\omega<0$) a phonon of frequency $\omega$.
To achieve ground state cooling $W_{op}\sim\mathcal{O}(\lambda^2/\Gamma_{op})$ must exceed $A_{op}^+$ \emph{and} the thermal heating rate $\gamma_r=\kappa N_{th}$,
where $\gamma_r\simeq k_BT/\hbar Q$ for relevant temperatures $ k_BT \gg\hbar \omega_r$.


The spectrum $S(\omega)$ is plotted in Fig.~\ref{fig:Cooling} for the sideband resolved regime $\lambda < \Gamma_{op} \ll  \Omega, |\Delta|, \omega_r$  where individual resonances can be assigned to transitions in the level diagram show in Fig.~\ref{fig:Cooling}(b).
Under resonance conditions, $\omega_{r}=\omega_{dg}$, cooling is dominated by transitions $ |n\rangle|g\rangle\rightarrow |n-1\rangle|d\rangle$ corresponding to the peak in the spectrum at $\omega \approx \omega_{r}$. This cooling process is partially compensated by heating transitions which can occur for non-zero populations $\rho_{ee}$, $\rho_{dd}$ of excited states $|e\rangle$ and $|d\rangle$. While $\rho_{ee}\!\sim\! 1/2$ under strong driving conditions, transitions $|n\rangle|e\rangle\rightarrow |n\!+\!1\rangle|d\rangle$ are detuned from resonance by $|\omega_{dg}-\omega_{ed}|=|\Delta|$ and do not significantly contribute to heating. The remaining resonant heating process, $ |n\rangle|d\rangle \rightarrow |n\!+\!1\rangle|g\rangle$, is proportional to the occupation of the dark state, which is populated only by optical dephasing processes.
Independent of $\Omega$ we obtain $\rho_{dd} \propto \alpha$ such that in the limit of ideal optical pumping $\alpha \rightarrow 0$, the dark state $|d\rangle$ remains unoccupied, thus enabling ground state cooling with a strongly driven spin.

\begin{figure}
\begin{centering}
\includegraphics[width=0.48\textwidth]{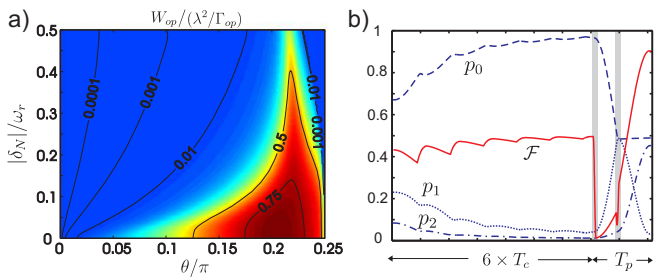}
\caption{(a) Optical cooling rate $W_{op}$ in the presence of a perturbation $H_{\rm nuc}=\hbar \delta_N S_z$ for $\alpha=0$ and $\Gamma_{op}/\omega_r=0.03$.  (b) Preparation of a superposition state $|\psi\rangle$ as discussed in the text where $\mathcal{F}={\rm Tr}\{ |\psi\rangle\langle \psi|\rho(t)\}$ and $p_i={\rm Tr}\{ |i\rangle\langle i|\rho(t)\}$. The sequence consists of 6 cooling cycles of duration $T_c\simeq\pi/4\lambda_g+2.5/\Gamma_{op}$ followed by a state preparation stage
with $T_p\simeq\pi/(4\lambda_g) + \pi/(\sqrt{8}\lambda_g)$. Gray bars indicate short $\pi/2$-rotations between states $|g\rangle$ and $|d\rangle$. The parameters used for this plot are  $\gamma_r/\lambda_g=0.01$, $\lambda_g/\omega_r=0.01$ and $\Gamma_{op}/\omega_r=0.05$.
} \label{fig:coolingrates}
\end{centering}
\end{figure}

We derive analytic expressions for $S(\omega)$ using the quantum regression theorem. To ensure resonance conditions we choose values for $\Delta<0$ and $\Omega$ which fulfill $\sqrt{\Delta^2+\Omega^2}=\omega_r/\cos^2(\theta)$ and study cooling as a function of the remaining free parameter $\theta\in [0,\pi/4]$.
For $\Gamma_{op}\ll \omega_{r}$ we then obtain $W_{op}=(\lambda^2/\Gamma_{op}) \mathcal{W}$ where
\begin{equation}
\mathcal{W}=\frac{  8\cos^4(\theta)\sin^2(\theta)}{[(1\!+\!\alpha)\!+\!\sin^2(\theta)][ 1\! +\!\cos^2(2\theta)\! +\! 3\alpha \sin^2(2\theta)/4]}.
\end{equation}
This function is maximized for $\theta\approx 0.2 \,\pi$ and for $\alpha\rightarrow 0$ we obtain an optimized damping rate of $W_{op}\simeq 0.8 \times \lambda^2/\Gamma_{op}$. Numerical values for $W_{op}(\theta)$ in the presence of hyperfine interactions are shown in Fig.~\ref{fig:coolingrates}(a). Here we find that close to $\theta\approx \theta_0$ not only we optimize $W_{op}$, but cooling is also most insensitive to perturbations. For given $W_{op} \gg \kappa$ the final occupation number is
\begin{equation}\label{eq:n0}
\langle n\rangle_0 \simeq   \frac{\alpha}{2} \tan^2(\theta) + \frac{1}{Q} \frac{k_BT}{\hbar W_{op}}\,.
\end{equation}
By choosing $\Gamma_{op}\sim \lambda$  and $\alpha\approx 0$ we find that conditions for ground state cooling coincide with the strong coupling regime.
Therefore, a single electronic spin can be used to optically cool the resonator into the quantum regime starting from initial temperatures $T\!\sim 0.1\!-\!1$ K. For $\alpha\neq 0$ the  apparent intrinsic limit for the present cooling scheme, $\langle n_0\rangle\sim \alpha/2$, can be overcome by employing a pulsed pumping strategy as discussed below.

Once prepared near its ground state, the resonator state can be completely controlled via Hamiltonian~\eqref{eq:HCavityQED}. The key mechanism that enables such
a control involves the unitary evolution under the resonant  JC Hamiltonian~\eqref{eq:HCavityQED} when $\omega_r = \omega_{dg}$. Specifically, for a time $t = \pi/2\lambda_g$ this evolution maps arbitrary spin states onto superpositions of motional states with zero and one phonon.
%
%
This procedure can be generalized for the generation of arbitrary states of the form  $|\psi\rangle=\sum_{n=0}^{M} c_n|n\rangle|g\rangle$ as proposed by
Law and Eberly~\cite{LawPRL1996}.
The basic idea is that we can construct a unitary transformation $U$ such that $U|\psi\rangle = |0\rangle$. Specifically,  $|\psi\rangle$ can be coherently mapped to a state $|\psi'\rangle$ with the maximal phonon number reduced by one after a free evolution $U_{JC}(\tau)=\exp(-iH_S\tau/\hbar)$ followed by a unitary rotation $U_x(\beta)=\exp(-i (\beta |d\rangle\langle g|+ \beta^* |g\rangle\langle d|))$, which can be implemented by an additional external microwave pulse. For appropriately chosen parameters $\tau$ and $\beta$ this combination removes first all population from state $|M\rangle|g\rangle$ and successively from state $|M\!-\!1,d\rangle$. By iterating this procedure one can step by step construct a unitary evolution $U$ which maps $|\psi\rangle$ onto the ground state $|0,g\rangle$. Then, by starting from $|0,g\rangle$ the inverse operation $U^{-1}$ will generate the target state $|\psi\rangle=U^{-1}|0,g\rangle$. For a given $M$ the state $|\psi\rangle$ can be generated within a time $t_M \leq M/2\lambda_g$ and a fidelity $\mathcal{F}\simeq (1- M \pi\gamma_r/2\lambda_g)\times p_0$, where $p_0$ is the initial occupation of state $|0,g\rangle$.


As an example of this procedure, we consider the generation of the state $|\psi\rangle=(|0\rangle - |2\rangle)|g\rangle/\sqrt{2}$ starting from a pre-cooled thermal state with $\langle n\rangle=0.5$. To prepare the initial state $|0,g\rangle$ with high fidelity, independent of $\alpha$, we use a pulsed pumping scheme. First, with $\Omega(t)=0$ the spin is optically pumped into the bare spin state $|0\rangle$. In a second step $\Omega(t)$ is turned on adiabatically such that the spin is prepared in state $|g\rangle$ while $|d\rangle$ and $|e\rangle$ remain unoccupied. Finally, for a time $t_{int}=\pi/(4\lambda_g)$ the system undergoes an  oscillation between states $|n\rangle|g\rangle$ and $|n- 1\rangle|d\rangle$. The repetition of this pulse sequence successively removes motional excitations and prepares the resonator in the state $|0,g\rangle$ with a probability $p_{0}\simeq 1- (\pi\gamma_r/4\lambda_g)$. Next, a sequence of two mw pulses and two partial swaps is used to prepare the cat-like state $(|0\rangle - |2\rangle)/\sqrt{2}$. Fig.~\ref{fig:coolingrates} shows the results of a numerical integration of master equation~\eqref{eq:MasterEquation} simulating 6 cooling cycles followed by the state preparation sequence. This example demonstrates that quantum ``engineering"
of motional states is possible using the present technique.  Finally,
the mapping procedure can be used for spin-mediated readout
of  the mechanical motional states.

In summary, we have shown that a single
electronic spin qubit in diamond can be strongly coupled to
the motion of a nano-mechanical resonator. Such a strong coupling
enables ground state cooling and quantum-by-quantum generation of
arbitrary states of the resonator mode. Potential applications
include the use of non-classical motional states for improved AFM-based force and magnetic sensing techniques~\cite{SidlesRMP1995}, as well as for tests of fundamental theories~\cite{MarshallPRL2003}.

\begin{acknowledgments}
We thank  M. Aspelmeyer, P. Zoller  and W. Zwerger for stimulating discussions.
This work was supported by ITAMP, the NSF and the Packard Foundation.
\end{acknowledgments}

\end{document}